\documentclass[a4paper,twocolumn,11pt,accepted=2024-11-16]{quantumarticle}
\pdfoutput=1

\usepackage{graphicx}%
\usepackage{dcolumn}%
\usepackage{bm}%
\usepackage[utf8]{inputenc} %
\usepackage[T1]{fontenc}    %
\usepackage{hyperref}       %
\usepackage{url}            %
\usepackage{booktabs}       %
\usepackage{amsfonts}       %
\usepackage{microtype}      %
\usepackage{amssymb,amsmath,amsthm,graphicx}
\usepackage{color}
\usepackage{comment}
\usepackage[ruled,vlined]{algorithm2e}
\usepackage{multirow}
\usepackage[caption=false]{subfig}
\usepackage{tabularx}
\usepackage{float}
\usepackage[numbers,sort&compress]{natbib}

\begin{document}


\title{
Learning to rank quantum circuits for hardware-optimized performance enhancement
}%

\author{Gavin S. Hartnett}
\orcid{0000-0002-6814-1809}
\email{gavin.hartnett@q-ctrl.com}
\author{Aaron Barbosa}
\orcid{0009-0009-3940-9886}
\author{Pranav S. Mundada}
\orcid{0000-0003-3316-0724}
\author{Michael Hush}
\orcid{0000-0002-6464-9234}
\author{Michael J. Biercuk}
\orcid{0000-0002-4371-2580}
\author{Yuval Baum}
\orcid{0000-0003-4631-8551}
\affiliation{Q-CTRL, Sydney, NSW Australia, and Los Angeles, CA USA}

\begin{abstract}
We introduce and experimentally test a machine-learning-based method for ranking logically equivalent quantum circuits based on expected performance estimates derived from a training procedure conducted on real hardware. We apply our method to the problem of layout selection, in which abstracted qubits are assigned to physical qubits on a given device. Circuit measurements performed on IBM hardware indicate that the maximum and median fidelities of logically equivalent layouts can differ by an order of magnitude. We introduce a circuit score used for ranking that is parameterized in terms of a physics-based, phenomenological error model whose parameters are fit by training a ranking-loss function over a measured dataset. The dataset consists of quantum circuits exhibiting a diversity of structures and executed on IBM hardware, allowing the model to incorporate the contextual nature of real device noise and errors without the need to perform an exponentially costly tomographic protocol. We perform model training and execution on the 16-qubit \textit{ibmq\_guadalupe} device and compare our method to two common approaches: random layout selection and a publicly available baseline called Mapomatic. Our model consistently outperforms both approaches, predicting layouts that exhibit lower noise and higher performance. In particular, we find that our best model leads to a $1.8\times$ reduction in selection error when compared to the baseline approach and a $3.2\times$ reduction when compared to random selection. Beyond delivering a new form of predictive quantum characterization, verification, and validation, our results reveal the specific way in which context-dependent and coherent gate errors appear to dominate the divergence from performance estimates extrapolated from simple proxy measures.
\end{abstract}
\maketitle

\section{Introduction}
A wide range of quantum computers have become available through  commercial, academic, and other public-sector platforms \cite{IBM_website, Rigetti_website, Amazon_website}. 
Both researchers and end-users are faced with a choice of architectures and qubit technologies, requiring a challenging process of selection in which the relative performance advantages and disadvantages of each specific platform must be assessed, synthesized, and used to choose a target hardware system for algorithmic execution.
As both the number and size of available quantum devices grow, the question of how to select the best device or sub-system of a device for the execution of a given quantum circuit is becoming increasingly important. Ultimately, this process is an ideal target for automation and abstraction if suitable tools for predicting the expected performance of a target algorithm on a hardware backend can be devised.

The topic of quantum computer benchmarking, or Quantum Characterization, Validation, and Verification (QCVV), aims to provide insights informing this selection process.
This field has generally focused on enabling the identification of quantum processors with the best overall performance characteristics based on abstracted and human-interpretable proxy measures describing low-level physical parameters such as qubit coherence or quantum logic gate errors, as derived from measurement protocols such as Randomized or Cross-Entropy Benchmarking \cite{emerson2005scalable, knill2008randomized, boixo2018characterizing, neill2018blueprint}.  
While it is intuitive to link low-level hardware characteristics to the device selection process, much subtlety is masked by the fact that the dominant framework for QCVV fundamentally trades between simplicity/efficiency and predictive power \cite{proctor2017randomized, proctor2022measuring, mavadia_QCVV}.  

\begin{figure*}[htp]
    \centering
    \includegraphics[width=1\textwidth]{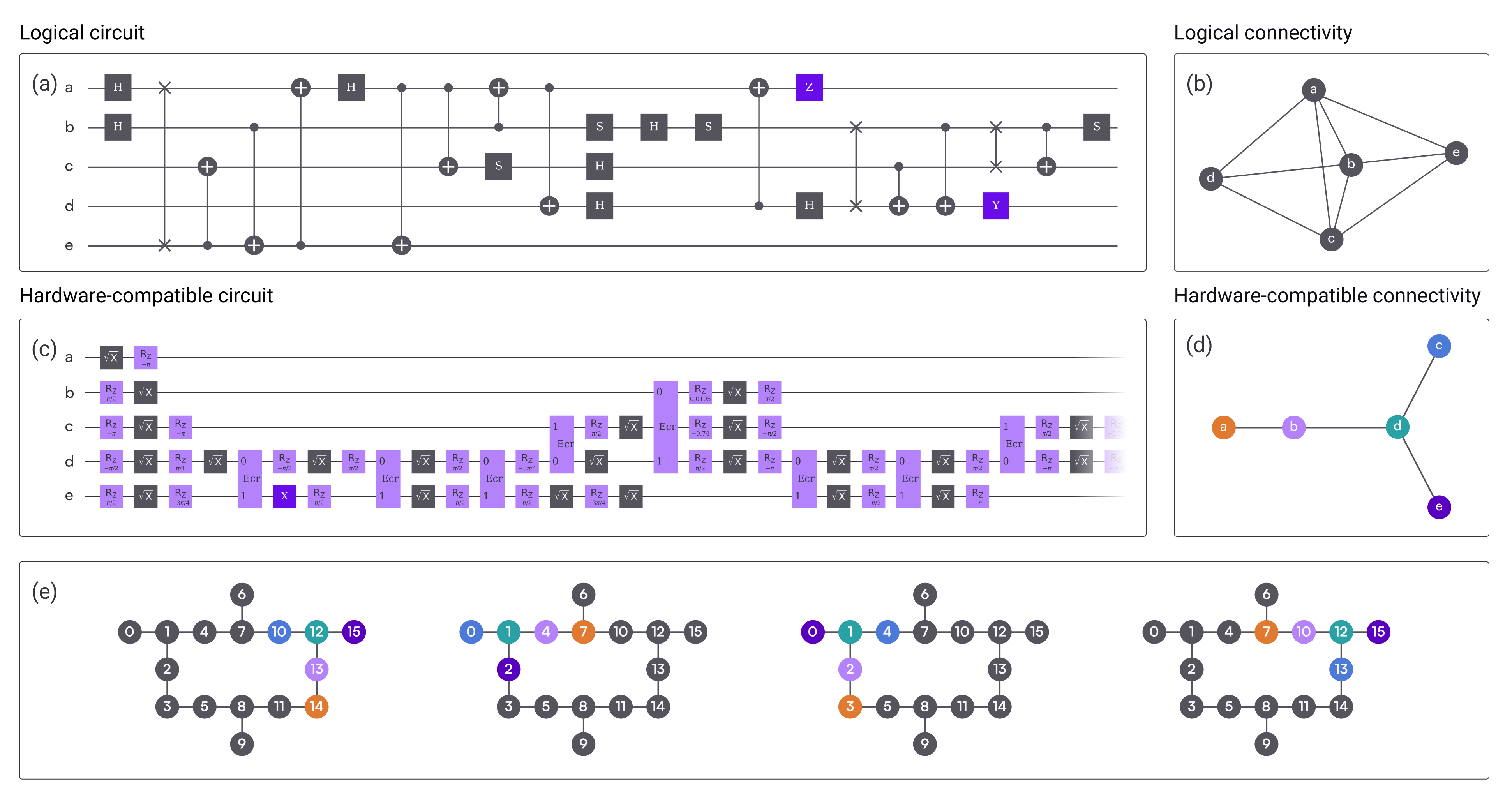}
    \caption{
    \label{fig:marquee_figure}
    Circuit transpilation and the task of layout selection. 
    (a) A representative initial circuit. Such circuits are unconstrained by hardware considerations, such as matching the device topology or consisting only of native gates.
    (b) The qubit connectivity graph of the initial circuit. Nodes represent the qubits appearing in (a) and edges represent 2-qubit gates.
    (c) The hardware-compatible circuit. Hardware transpilation transforms the initial circuit into a unitarily equivalent, hardware-compatible circuit. Shown here is the hardware-transpiled circuit corresponding to the initial circuit on the \textit{ibmq\_guadalupe} device (only a portion is shown for brevity). 
    (d) The qubit connectivity graph of the hardware-transpiled circuit. Unlike the initial connectivity graph, the transpiled connectivity graph is a subgraph of the device topology (shown in (e)). The node labels correspond to the qubit labels in (d).
    (e) The layout is a map between circuit qubits and physical qubits. There are typically many such mappings; shown here is a subset of four layouts. The goal of layout selection is then to choose the best-performing layout for execution on the device.
    }
\end{figure*}

Extrapolating from hardware-level proxies to system-level performance is complicated by the fact that most common scalable QCVV approaches are by design insensitive to the myriad sub-structures, or contexts, that can appear within quantum circuits and which can have important impacts on the performance.   This is in part due to the historic use of these metrics in analyses of fault tolerance, in which errors are typically (and speciously) assumed to be statistically independent.
For example, randomized benchmarking (RB) is relatively efficient to implement and easy to interpret, yet it masks coherent errors exhibiting correlations in time over many gates in a circuit~\cite{mavadia_QCVV, carvalho_Gates}.
Similarly, crosstalk may change both the magnitude and the nature of the error of a specific gate depending on the existence or absence of simultaneous gates on neighboring qubits~\cite{carvalho_Gates, Edmunds_ErrorCorrelationsRB} - contexts that are ignored or averaged over in standard gate-level QCVV protocols.
In particular, errors that occur during idling periods are in general poorly captured by most standard methods, both at the single qubit level where $T_2$ effects are relevant, and at the multi-qubit level where quantum crosstalk can lead to an unwanted unitary evolution between neighboring qubits \cite{ZZ_suppression2020, mundada2023experimental, bronn2024multiqubitDD}. 
Furthermore, the overall effect of an error on the measured fidelity of a quantum circuit will be dependent on the structure of the circuit; 
random circuits will generally be less sensitive to over-rotation gate errors due to self-cancellations of the random excess angles~\cite{ball_RB}, whereas circuits with a repetitive structure, such as Trotter circuits, will be very sensitive to these types of errors \cite{proctor2017randomized, mavadia_QCVV, proctor2022scalable}.  
The aggregate effect of these various circumstances - all of which are routinely encountered in the execution of real algorithms on hardware - tends to cause QCVV protocols to only be loosely predictive of algorithmic performance~\cite{proctor2022measuring, mundada2023experimental}.  

Here, we focus on the concrete problem of \emph{layout selection} within a single processor: the mapping of virtual or abstract qubits in a compiled circuit to physical device qubits in a manner that maximizes circuit fidelity.  Previous work has examined the topic of layout selection via various methods~\cite{zhang2023characterizing, acampora2021deep, mao2023quantum, hothem2023learning, hothem2023predictive, peters2022qubit} (see also \cite{silver2024qompose, 
shaik2023optimal}), but has not presented a comprehensive ranking procedure based on measurements of real circuits executed on hardware.  This omission ignores the essential contextual information that is known to cause divergence between proxy measure predictions and achievable performance.

\begin{figure*}[htpb!]
    \centering
    \includegraphics[width=1.0\textwidth]{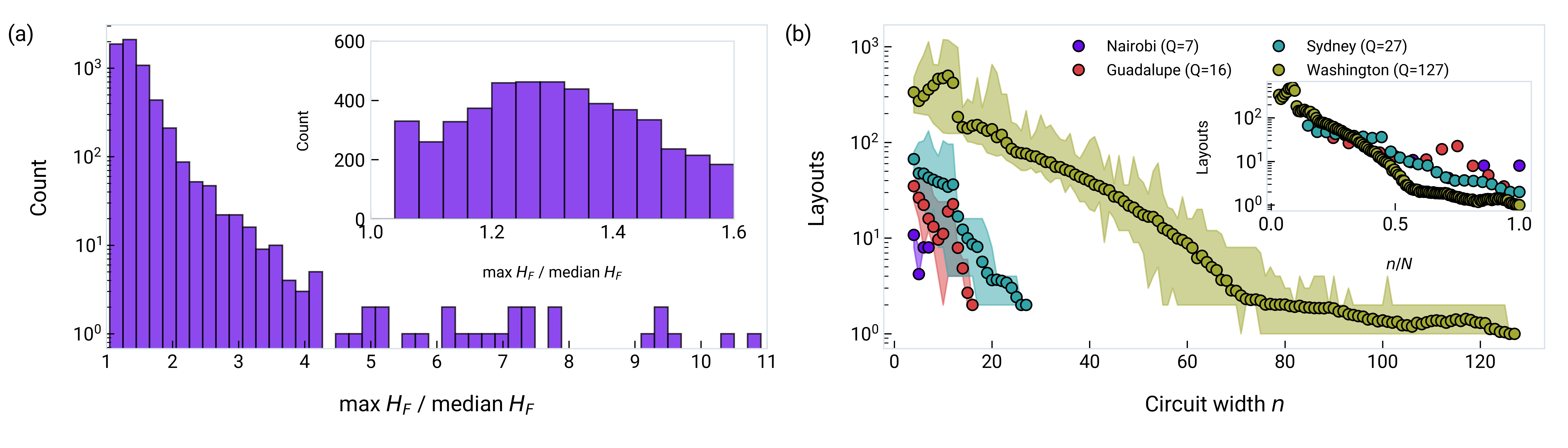}
    \caption{
    \label{fig:layouts}
    The importance of layout selection.
    (a) 
    Distribution of the performance quality ratio between the median- and maximum-performing layouts as measured by the Hellinger fidelity, $H_F$, evaluated over large a data set of more than 6000 circuits (see the main text for more details about the data set). 
    The inset shows a zoomed-in and higher-resolution plot for performance qualities between 0 and 60\%.
    (The first bar in the inset has zero counts, i.e., the difference between the top and median performing layouts exceeded $4\%$ for all circuits.)
    (b) The number of layouts, averaged over 500 random initial circuits, as a function of the circuit width $n$.
    The markers indicate averages, and the shaded region indicates the range.
    The inset shows the average number of layouts as a function of $n/N$, the fraction of device qubits used.
    Note that these results assume bi-directional device graphs.
    }
\end{figure*}

In this work, we develop a new form of \emph{predictive} QCVV for layout selection for arbitrary target circuits based on a heuristic, machine learning solution. We report direct measurements on an IBM quantum computer revealing that different layouts of individual target circuits exhibit up to a factor of $11\times$ variation in the ratio of the maximum achieved to median Hellinger fidelity, highlighting the importance of this task.   
We develop a procedure to rank the expected relative performance of logically equivalent layouts for arbitrary target circuits within a quantum processor, drawing from a large body of work in machine learning where this general problem is known as ``Learning to Rank'' \cite{liu2009learning, li2011short, burges2005learning, cao2007learning}.
In so doing, this method prioritizes the \textit{relative} performance between logically equivalent permutations of a user-defined target circuit over accurate estimates of absolute performance, and omits the reliance on human-interpretable proxy measures.   Our method trains a circuit-score model over a dataset of circuits exhibiting diverse structural characteristics, and parameterizes the target-circuit score in a physics-based phenomenological manner.  The model then assigns to each layout of the target circuit a numerical score used to quantify relative expected performance (a higher score corresponds to a higher expected performance).  We test the efficacy of the ranking procedure using direct fidelity measurements of the different layouts on hardware, and compare our learning-to-rank method against alternative layout-selection approaches.  We also compare five different candidate loss functions and quantitatively demonstrate that the listwise Rank MSE loss function performs best across multiple metrics. For the best model, measurements demonstrate a reduction in selection error -- the relative loss in Hellinger fidelity due to sub-optimal layout selection -- by $3.2\times$ ($1.8\times$) when compared to a random and a baseline error-informed layout selection strategy, respectively, validating the utility of this approach.  

The rest of the article is organized as follows. 
In Sec.~\ref{sec:layoutselection} we briefly review the problem of layout selection.
In Sec.~\ref{sec:circuitranking} we introduce our approach, and then perform experiments on real quantum hardware in Sec.~\ref{sec:results}, before concluding with a discussion in Sec.~\ref{sec:discussion}. 
Further technical details are included in Appendix~\ref{app:score} and \ref{app:lossfunctions}.

\section{The layout-selection problem \label{sec:layoutselection}}
Hardware transpilation takes a generic circuit as input and modifies it so that it consists only of operations native to a given quantum backend. 
Crucially, the circuit modifications made in this step must preserve the logical operation of the circuit, i.e., the unitary transformation effected by the circuit must remain unchanged (perhaps up to a specified non-zero tolerance).
For example, circuit identities involving SWAP gates may be used to adjust the qubit connectivity of a circuit to match a given device topology. 
Most backends natively support only single- and two-qubit gates, and throughout this work, we will assume this as well ($q$-qubit gates for $q > 2$ can be decomposed in terms of 1- and 2-qubit gates during the compilation step). 

The transpiled circuit and quantum backend device may each be associated with graphs that will be central to the formulation of the layout selection problem. 
For every two-qubit gate present in the circuit, the circuit graph will contain an edge connecting the participating qubits.
Similarly, the edges of the device graph will correspond to the qubit connectivity of the backend. The impact of this transition from abstract logical circuits to hardware-compatible circuits is illustrated in Fig.~\ref{fig:marquee_figure}(a)-(d).
Generally, the sparse connectivity of the device translates into an increased complexity of the transpiled circuit.

Further, a circuit will ``fit'' on a device if the circuit graph is subgraph-isomorphic to the device graph.
Such subgraph isomorphisms are represented as layouts, mappings from the abstract circuit qubits to specific physical device qubits, as shown in Fig.~\ref{fig:marquee_figure}(d)-(e). The set of all layouts for a given transpiled circuit can be efficiently identified using VF2, a well-known subgraph isomorphism algorithm \cite{cordella2004sub}, as well as related algorithms \cite{juttner2018vf2plusplus}. 

Once the set of allowed layouts for a target circuit has been enumerated, the problem of {\sl layout selection} is to identify a single layout to be used for the actual execution of the circuit on a device, mapping abstract or virtual qubits to physical devices and gate sets. 
This introduces a mechanism for significant performance variability, as gate errors and coherence times can exhibit order-of-magnitude variation across a single device~\cite{carvalho_Gates}. In addition, the fidelity and execution time of an individual gate will generally be context-dependent, varying with the selected subset of participating qubits. 

Due to the variability among layouts, layout selection is an integral sub-task in quantum circuit compilation that can contribute substantially to the overall performance of the circuit when executed on hardware. 
This is demonstrated in Fig.~\ref{fig:layouts} (a), which shows experimental measurements of the circuit fidelity achieved for over 6000 initial circuits. 
For each circuit, all possible layouts were identified and executed on the 16-qubit \textit{ibmq\_guadalupe} device. 
The figure shows the distribution of the ratio between the performance of the best and median layouts, as measured using the Hellinger fidelity. 
\footnote{
The Hellinger fidelity is a measure of the closeness of two probability distributions.
For distributions over the set of length $n$ bitstrings, 
${H_F(p, q) = \left( \sum_{x \in \{0,1\}^n} \sqrt{p(x) q(x)} \right)^2}$.
$H_F = 1$ corresponds to perfect agreement, and $H_F = 0$ corresponds to $p$ and $q$ having zero overlap. 
In this work, the two distributions of interest are $\hat{q}$, the empirical bitstring distribution obtained through bitstring counts in $N_{\text{shots}}$ measurements of an executed circuit, and $p$, the ideal, noise-free bitstring distribution.
We use the overloaded notation $H_F(C)$ to denote $H_F(p, \hat{q})$. 
}
All circuits are executed in an identical manner, and comparisons are made strictly between measured results vs theoretical predictions. Any deviation from the value one on the horizontal axis of Fig.~\ref{fig:layouts} (a) represents a deleterious impact on circuit fidelity as a result of sub-optimal layout selection.
For example, a value of $3$ indicates that with 50\% probability, a randomly selected layout will exhibit $3\times$ worse performance than that achieved by the top-performing layout.    For a majority of layouts,f the degradation in performance between the top and median layouts is substantial - for example,
about 28\% of circuits exhibit a max-to-median ratio that exceeds $1.5\times$. 

Layout selection is further complicated by the combinatorially large number of possible layouts - a number determined by the connectivity graphs of the transpiled circuit and the device. 
In general, the two key quantities determining the number of layouts are the circuit width, $n$, and the number of qubits in the device, $Q$.  For $n \ll Q$ the number of layouts can be extremely large, and it gradually diminishes as $n\to Q$, given there are fewer accessible ways to ``fit'' a large circuit on the device.
This relationship is demonstrated in Fig.~\ref{fig:layouts} (b), which shows the number of layouts for random circuits of varying widths compiled for the IBM devices \textit{ibm\_nairobi}, \textit{ibmq\_guadalupe}, \textit{ibmq\_sydney}, \textit{ibm\_washington}, which have $Q=7, 16, 27$, and $127$, respectively. 
For circuits using less than half of the available qubits on the 127-qubit \textit{ibm\_washington} device, the number of layouts can reach several hundred. 

Given the variability of circuit performance demonstrated in Fig.~\ref{fig:layouts} -- with the maximum circuit sometimes exhibiting more than $10\times$ higher performance than the median -- it becomes imperative to define a procedure to predict which among this potentially large selection of equivalent circuit layouts is most likely to exhibit the highest performance.  To address this, in the following, we present a learning algorithm allowing for approximation of this ordinal ranking that can be run prior to compilation and execution on hardware.

\section{
Circuit ranking for layout selection \label{sec:circuitranking}
}

\begin{figure*}[ht]
    \centering
    \includegraphics[width=1.0\textwidth]{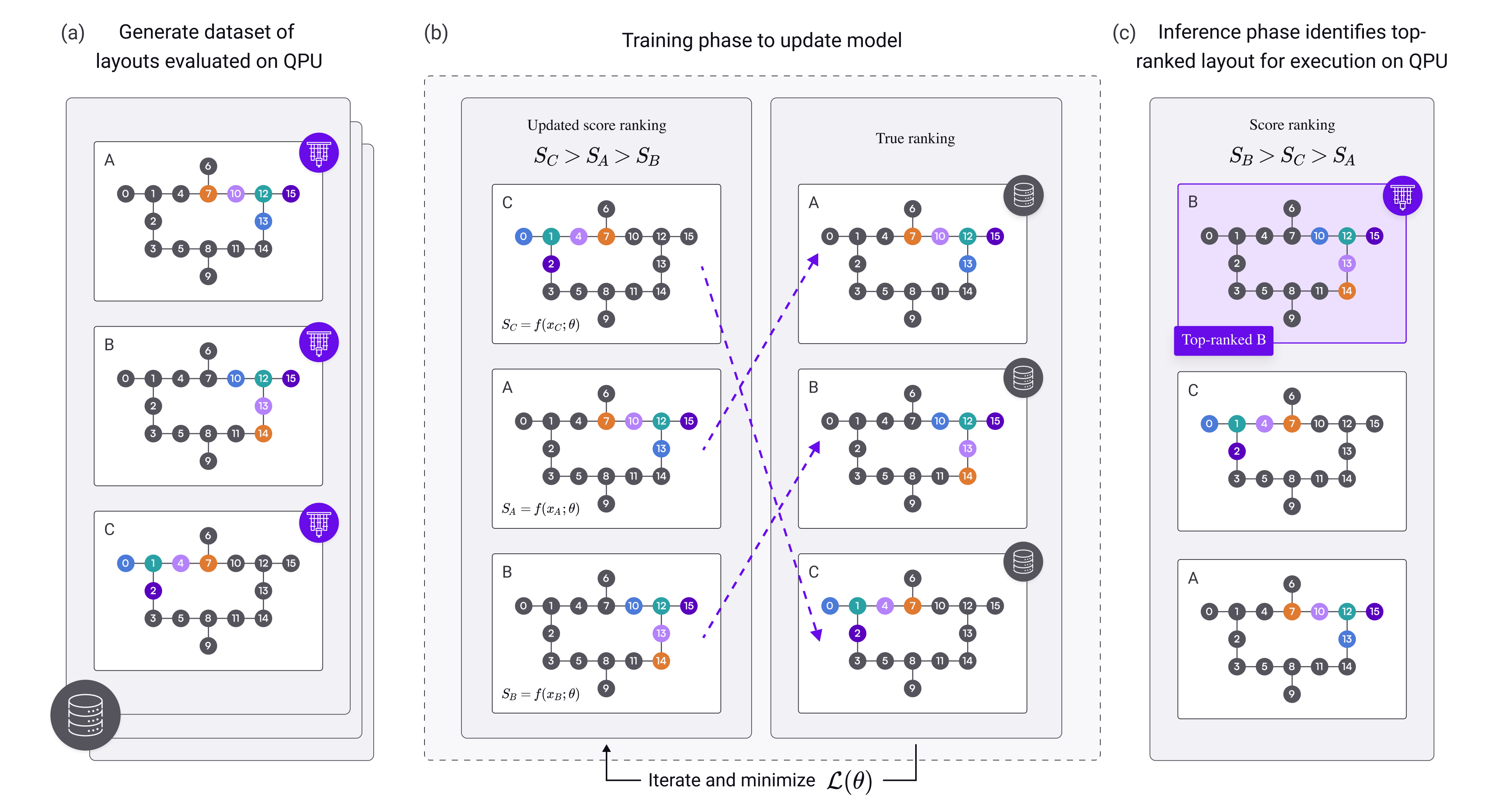}
    \caption{
    \label{fig:score-ranking}
      Machine learning pipeline for layout selection. (a) Layout generation. Graphs indicate connectivity of \textit{ibmq\_guadalupe} device:  graph nodes indicate qubits and colored nodes are guides to the eye indicating different layouts for a notional five-qubit circuit. Only three layouts, A-C, shown for visual clarity. Different layers represent sets of layouts for different initial circuits.  Purple QPU icons indicate hardware execution.  The black database icon indicates that data from hardware execution form a dataset.
      (b) Overview of training phase.  Ordering of layouts is now set by ranking conducted using current state of model.  Purple dashed arrows indicate mismatches between model prediction and true ranking determined by database.  The lower arrow indicates an iterative loop to minimize loss function via model parameter updates (see Table~\ref{table:lossfunctions}).
      (c) Overview of inference phase. The highest-ranking layout is selected for hardware execution, indicated by purple shading. There is no restriction that these layouts be present in the dataset, and in the primary use-case, the goal will be to predict layout rankings when the true ranking is unavailable
      }
\end{figure*}

We now present an overview of a practical and efficient machine-learning approach for ranking-based layout selection. We use a score-based approach for ranking circuit performance.
The circuit score $S$ is a function that maps scheduled circuits to a real-valued numerical score: higher scores correspond to higher estimated fidelities. 
In a good ranking procedure, ordering layouts by circuit score should lead to the same ordinal ranking among circuits as a chosen performance metric derived from measurements on hardware.  In this work, we take this measured quantity to be the Hellinger fidelity $H_F$ between the ideal bitstring distribution and the empirically observed bitstring distribution. 

Therefore, given two circuits $C_1, C_2$, the following relation should hold:
\begin{equation}
    \label{eq:ideal_ranking}
    H_F(C_1) \lesssim H_F(C_2) \quad \text{iff} \quad S(C_1) \lesssim S(C_2) \,.
\end{equation}
Our objective is now to learn score functions that approximately obey Eq.~\ref{eq:ideal_ranking}. 

The circuit score is modeled using a hybrid approach that augments phenomenological physical models of device errors with an experimental-data-driven, machine learning approach. 
First, we assume that the score may be written as a product of terms 
\begin{equation}
    S = \prod_{a} (S_a)^{p_a} \,,
\end{equation}
with each $S_a \in [0,1]$. 
The power parameters account for the fact that the circuit structure affects the relative importance of different noise mechanisms; for example, deep random circuits will be more sensitive to incoherent errors and less to systematic gate calibration errors compared to shallow structured circuits (such as Trotter circuits). Complete definitions of the $S_a$ scores are presented in Appendix~\ref{app:score}.

These scores are parameterized in terms of physically meaningful quantities that are typically estimated by the hardware provider, such as gate and measurement error rates, $T_1$ times, and two-qubit interaction couplings.  
However, standard calibration routines do not provide generalized context-aware information: during calibration specific circuit structures are deliberately employed to ``amplify'' the physical process being characterized.  
This intentional bias in characterization contributes to errors when ranking layouts, by missing the link between the context of the circuit to be executed and the isolated performance of the underlying hardware. 

We address this shortcoming by promoting these parameters to be learnable, and the circuit score to be a machine learning model. 
The parameters are fit by minimizing a loss function $\mathcal{L}(\theta)$ over a training dataset, where we introduce $\theta$ as the vector of model parameters: 
\begin{equation}
    \theta_* = \text{argmin}_{\theta} \, \mathcal{L}(\theta) \,.
\end{equation}
\noindent The loss function provides a continuous and differentiable way to penalize deviations from Eq.~\ref{eq:ideal_ranking}. 
We consider multiple loss functions, summarized in Table~\ref{table:lossfunctions}, reflecting the fact that there does not appear to be a clear preferred choice in the Learning-to-Rank literature. 
These can be grouped into three categories: pointwise, pairwise, and listwise. 
Pointwise approaches do not involve any comparisons, pairwise approaches involve comparisons between pairs of layouts, and listwise approaches involve comparisons across a set of all possible layouts for a given initial circuit. One important consideration of the pairwise loss, compared to the pointwise MSE loss, is that it facilitates a quadratic expansion of the dataset: an initial dataset of $N$ distinct circuits yields $N(N-1)/2$ distinct pairs of circuits. This expansion can be useful because it allows for many more training examples to be drawn from a given dataset, but it also has drawbacks because the dataset size can become unwieldy as well as depart from being independent and identically distributed (iid) \cite{cao2007learning}. For these reasons, we did not implement a pairwise loss in our experiments. Details for the various loss functions considered are presented in Appendix~\ref{app:lossfunctions}. 

\begin{table*}[ht]
\centering
\caption{\label{table:lossfunctions} Learning to rank loss functions. Multiple loss functions were considered; each penalize deviations from a perfect ranking in a different manner. The Score MSE loss encourages the score function to agree pointwise with the Hellinger fidelity. The Pearson and Soft Spearman losses encourage the score to be perfectly correlated with Hellinger fidelity across a batch of layouts. The Rank MSE encourages the Hellinger fidelity rank to agree with the score rank across a batch of layouts. Lastly, the Negative Log-Likelihood (NLL) loss maximizes the probability of predicting the top-$K$ layouts, with $K$ a hyper-parameter.}

\vspace*{2mm}
\fontsize{9}{9}\selectfont
\begin{tabular}{p{0.2\linewidth}p{0.1\linewidth}p{0.6\linewidth}}
    \multicolumn{1}{l}{Loss function} & \multicolumn{1}{l}{Type} & \multicolumn{1}{l}{Optimization description} \\[0.1cm] \hline \hline
    \multicolumn{1}{l}{Score MSE} & pointwise & Minimize difference between score and Hellinger fidelity \\[0.1cm] %
    \multicolumn{1}{l}{Pearson} & listwise & Maximize correlation between score and Hellinger fidelity \\[0.1cm] %
    \multicolumn{1}{l}{Soft Spearman} & listwise & Maximize correlation between score and Hellinger fidelity \\[0.1cm] %
    \multicolumn{1}{l}{Rank MSE} & listwise & Minimize difference of true and predicted rank \\[0.1cm] %
    \multicolumn{1}{l}{NLL} & listwise & Maximize model probability of observed top-$K$ layouts \\[0.1cm] %
\end{tabular}
\end{table*}

For Learning-to-Rank problems, a dataset of objects and their true ranking is required.
In the present context, the objects are transpiled circuits differing only in their layout, and the true ranking is provided by the Hellinger fidelity deduced through hardware execution.
Such a dataset can be created by first sampling an initial circuit from a statistical ensemble (to be discussed presently), transpiling that circuit into multiple circuits differing only in their layout, executing each transpiled circuit on a quantum device, and finally computing the Hellinger fidelity. This procedure results in a dataset of the form
\begin{equation}
    \mathcal{D} = \{ \mathcal{C}_i, \mathcal{F}_i \}_{i = 1, ..., N_B} \,,
\end{equation}
where $\mathcal{C}_i = \{C_{i,1}, C_{i,2}, ... C_{i,L_i}\}$ is a batch of $L_i$ circuits, one for each possible layout, and where ${\mathcal{F}_i = \{ H_F(C_{i,1}), H_F(C_{i,2}), ..., H_F(C_{i,L_i}) \}}$ is the collection of Hellinger fidelities for each circuit in the batch. 
Here, $N_B$ counts the total number of layout-batches considered, which is equal to the number of initial, un-transpiled circuits. 
Note that the Hellinger fidelity compares the empirically measured and ideal bitstring distributions, and therefore its computation presupposes the ability to simulate the circuit. 

To ensure that our model leads to an accurate ranking across a range of quantum algorithms, the circuit score is learned by minimizing the loss function averaged over a training dataset consisting of a diversity of circuits. These are chosen to reflect a wide range of structures and depths commonly encountered in real algorithms, making them differentially sensitive to different types of error:
\begin{itemize}
    \item Bernstein-Vazirani~\cite{bernstein1993quantum} circuits have a shallow, sequential structure, and are relatively more sensitive to idling and SPAM errors. 
    \item Inverse QFT circuits consist of a layer of single-qubit gates, followed by the inverse quantum Fourier transform circuit primitive. The initial layer is such that the ideal output state is a computational basis state.  These are deeper circuits (quadratic in $n$), and are more sensitive to 2-qubit gate errors.  
    \item Clifford-conjugated-Pauli circuits (also referred to as Mirror circuits in \cite{proctor2022measuring, proctor2022scalable}) are of the form ${C = (C_{\text{Clifford}})^{\dagger} \left( \otimes_{k=1}^n P_k \right) C_{\text{Clifford}}}$, where $P_k$ is a single-qubit Pauli gate, and $C_{\text{Clifford}} \in \text{Cl}(n)$ is a circuit implementing a random Clifford element (either chosen uniformly at random using the algorithm of \cite{bravyi2021hadamard} or by uniformly sampling a pre-determined number of Clifford gates). These have a high density with randomized gates and are therefore mainly sensitive to incoherent errors. 
    \item QAOA circuits~\cite{farhi2014quantum} exhibit arbitrary rotations and are therefore mainly sensitive to over-rotation coherent errors.
\end{itemize}
In addition to containing a broad range of contexts, these circuits were chosen because they are either efficiently simulable or can be easily constructed to return a desired output, which allows the Hellinger fidelity between the ideal bitstring distribution and the empirically observed distribution to be easily calculated. 
While not efficiently simulable, QAOA circuits are included as they are currently a very popular algorithm for execution on existing quantum devices. 

A summary of the machine learning pipeline for layout selection is presented in Fig.~\ref{fig:score-ranking}. To train the ML model, a dataset of circuits is first generated. This is illustrated in Fig.~\ref{fig:score-ranking}(a), where each layer represents a batch of layouts transpiled to the hardware for a different circuit. Each layout is then executed on the quantum processing unit (QPU) and the Hellinger fidelity is computed for all layouts by comparing the ideal bitstring distribution against the empirical distribution. Together, these batches of executed layouts and their fidelities form a dataset of labeled data. In the training phase, illustrated in Fig.~\ref{fig:score-ranking}(b), the batches of layouts in the dataset are ranked and the model iteratively updated. For each batch, the current state of the model produces a score ranking that is compared against the true ranking derived from the Hellinger fidelity. Mismatches in the two rankings are penalized by the differentiable loss $\mathcal{L}(\theta)$. The model parameters $\theta$ are updated through multiple iterations of a gradient-based algorithm.
In the inference phase, illustrated in Fig.~\ref{fig:score-ranking}(c), the goal is to use the trained model to predict a single layout for execution on the QPU. The input is a single transpiled circuit and the most up-to-date backend information (such as gate errors and $T_1$ times). The set of all valid layouts is generated using the subgraph isomorphism capability of the Mapomatic package, and a score $S$ is assigned to each layout by the already trained model. The layouts are ranked according to the circuit score, and the circuit with the largest score is chosen for execution on the device.

\section{Experimental results \label{sec:results}}

\begin{figure*}[ht]
    \centering
    \includegraphics[width=0.8\textwidth]{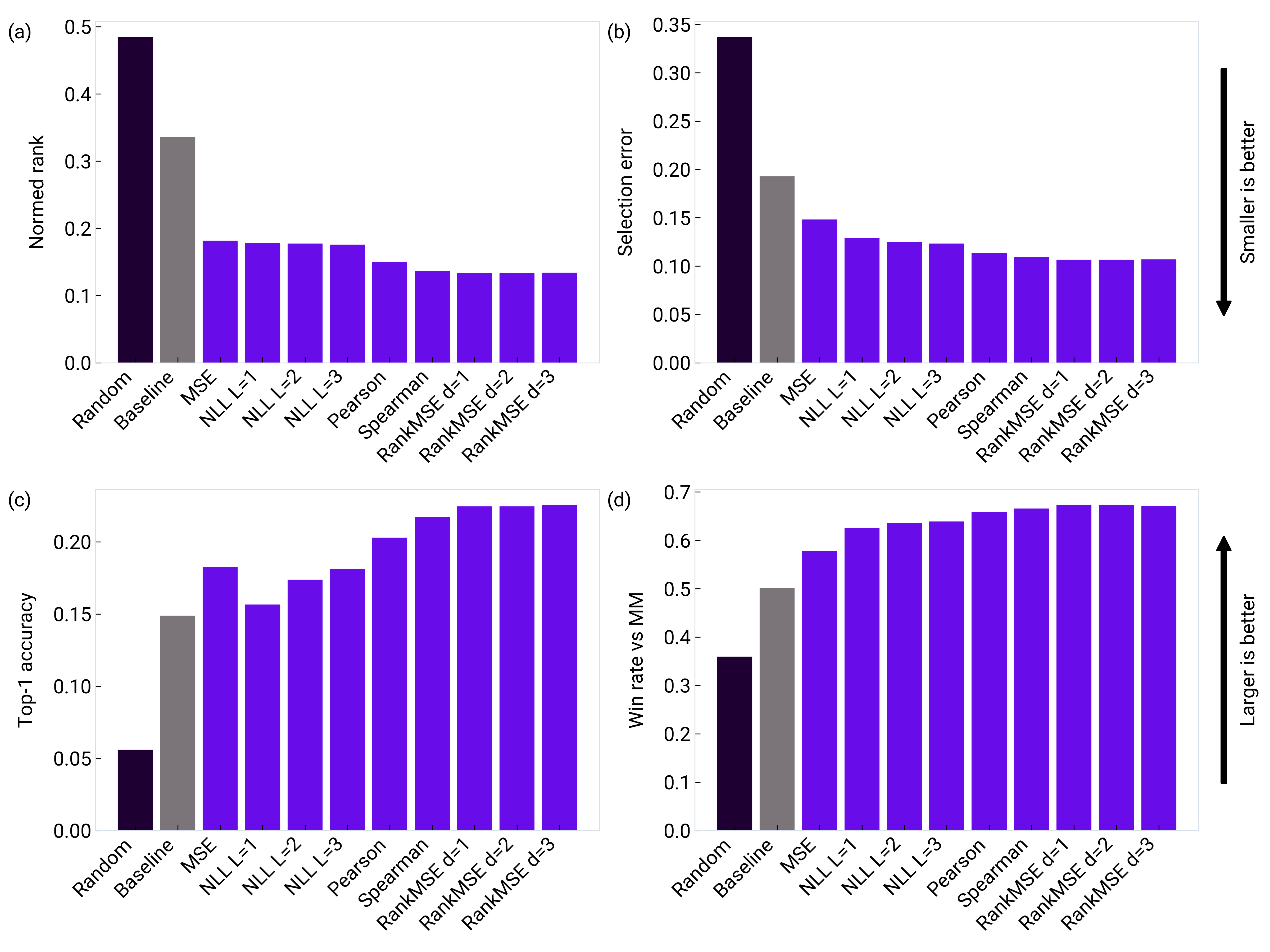}
    \caption{
    Model performance. The performance of circuit score models trained on the loss functions listed in Table~\ref{table:lossfunctions}, as measured by multiple evaluation metrics averaged over the test set.
    The win rate is measured against Mapomatic (MM).
    Note that a lower value is better for the normed rank (a) and selection error (b), whereas a higher value is better for the top-1 accuracy (c) and win rate (d). 
    To account for statistical variations in the optimization and in the training/testing split, for each loss function 10 random seeds are used, and numerical values are averaged over these. 
    }
    \label{fig:results}
\end{figure*}

We perform circuit ranking using training data and experimental measurements executed on IBM hardware. 
In parameterizing the circuit score, we therefore consider four common error mechanisms for fixed-frequency superconducting qubit devices: gate error (for both one- and two-qubit gates), measurement error, $T_1$ error, and $ZZ$ crosstalk \cite{mundada2023experimental}. Each error type is separately modeled using a parameterized phenomenological approach based on device parameters resulting in a set of four scores: $S_{\text{gate}}$, $S_{\text{msmt}}$, $S_{T_1}$, and $S_{ZZ}$. 

A training dataset of different circuit types is generated for the 16-qubit \textit{ibmq\_guadalupe} device, and split into training and testing sets according to an 80:20 ratio. 
These 6020 initial circuits become 132,866 circuits after hardware transpilation, and the VF2 algorithm (as implemented in \cite{nation2023suppressing}) is used to calculate all possible layouts. 
For each circuit, we compute the ideal bitstring distribution and execute the circuit on the quantum computer using $N_{\text{shots}} = 4096$ in order to obtain the Hellinger fidelity.
To ensure a clear learning signal for our circuit score, we apply an automated error suppression pipeline from our software suite to each executed circuit \cite{mundada2023experimental} to maximize execution fidelity. 
Due to the large number of layouts considered, this data collection was spaced over a period of 9 months (from Nov 2022 to July 2023).  
All models are trained using the gradient-based Adam optimizer \cite{kingma2014adam} and the OneCycleLR learning rate scheduler \cite{smith2019super}.

We benchmark our approach against both random selection and selection via the Mapomatic package \cite{nation2023suppressing}. Random selection is the default approach when utilizing a standard compiler.  
Many informed users employ Mapomatic, a layout selection method that estimates circuit fidelity by aggregating the backend-estimated fidelities of each individual operation (i.e., gate errors, measurement errors, etc). 
\footnote{
	Note that Mapomatic is the name of both the layout selection algorithm as well as the Python package that first identifies all valid layouts and then ranks them according to a simple heuristic score - here, we will use Mapomatic to specifically refer to the score-based layout selection algorithm.
	}
Mapomatic also uses a scoring function to rank layouts; the key points of difference are in the choice of score function and the fact that in the models developed here, the parameters are \emph{learned} by training the model over a dataset containing a diverse range of circuit contexts.

To evaluate different methods, we quantify the quality of a layout selection procedure using a number of evaluative metrics. 
First, the true rank of the selected layout 
serves as the natural metric for this learning-to-rank problem.
We adopt the convention that the true rank $R$ ranges from 1 (top-ranked) to $L$ (worst-ranked) for a batch of $L$ layouts. 
As $L$ can vary significantly depending on the algorithm and device, it is also useful to consider a normed rank $r = (R-1)/(L-1)$, with $r \in [0,1]$ (a lower value is better).
Second, the Top-1 accuracy is defined as the fraction of circuits for which the model predicts the top-ranked layout.
Third, the selection error, defined as 
${1 - H_F(C_{\text{pred}}) / H_F(C_{R=1})}$,
measures the fraction of the optimal Hellinger fidelity that is lost by choosing a sub-optimal layout.
In particular, the top-ranked layout by definition has a selection error of zero.

\begin{table*}[ht]
\caption{
Model performance. The performance of the circuit score model is measured by multiple evaluative metrics. Median rank $r$ corresponds to the median normed rank, Median rank $R$ corresponds to the median un-normalized rank, Selection Err corresponds to the selection error, and Top-1 Acc corresponds to the Top-1 accuracy. Each row corresponds to a circuit score model trained on a different loss function, averaged over 10 random seeds. Uncertainties correspond to 1 standard error. For comparison, the results for random layout selection and Mapomatic are also shown. The win rate of each method against Mapomatic (MM) is shown in the final column.
}
\label{table:results}
\fontsize{9}{9}\selectfont
\begin{center}
\begin{tabular*}{\textwidth}{l@{\extracolsep{\fill}}l@{\extracolsep{\fill}}ccccc}
\toprule
Model & Loss & Median rank $r$ & Median rank $R$ & Selection Err (\%) & Top-1 Acc (\%) & Win Rate (\%) \\
\hline \hline
Random \, & N/A & 0.4838(57) & 10.00(15) & 33.64(30) & 5.56(27) & 35.86(42) \\ \hline
Mapomatic \, & N/A & 0.3351(53) & 7 & 19.24(11)  & 14.86(30) & 50 \\
\hline
Circuit Score \, & MSE & 0.1808(22) & 4.50(17) & 14.759(92) & 18.23(26) & 57.71(32) \\ \cline{2-7}
 & NLL$_{K=1}$ & 0.17701(53) & 4.20(13) & 12.818(80) & 15.63(32) & 62.48(36) \\ \cline{2-7} 
 & NLL$_{K=2}$ & 0.1764(35) & 4.20(13) & 12.44(24) & 17.34(41) & 63.40(52) \\ \cline{2-7}
 & NLL$_{K=3}$ & 0.17494(42) & 4.10(10) & 12.29(11) & 18.09(38) & 63.80(34) \\ \cline{2-7}
 & Pearson & 0.1487(50) & 4.0 & 11.295(62) & 20.26(31) & 65.73(25) \\ \cline{2-7}
 & Spearman & 0.1358(23) & 3.90(10) & 10.863(68) & 21.67(46) & 66.47(27)  \\ \cline{2-7}
 & Rank MSE$_{d=1}$ & \bf{ 0.13275(39)} & \bf{ 3.70(15) } & \bf{10.603(97)} & 22.41(41) & \bf{67.19(18)} \\ \cline{2-7}
 & Rank MSE$_{d=2}$ & \bf{0.13275(39)} & \bf{ 3.70(15) } & \bf{10.603(97)} & 22.41(41) & \bf{67.19(18)} \\ \cline{2-7}
 & Rank MSE$_{d=3}$ & 0.13304(29) & 3.80(13) & 10.638(78) & \bf{22.52(46)} & 67.03(18) \\
\bottomrule
\end{tabular*}
\end{center}
\end{table*}

Fig.~\ref{fig:results} presents the key results of our machine-learned circuit score method for layout selection, comparing the various metrics and loss functions introduced above; numerical values are tabulated in Table~\ref{table:results}. 
In all cases, the circuit score outperforms the baseline provided by Mapomatic or random selection. The best-performing model across all metrics were trained on the Rank MSE loss function.  
Compared to random selection, the Rank MSE models correspond to a 3.6$\times$ improvement in the normed rank and a 3.2$\times$ improvement with respect to the selection error. 
Compared to Mapomatic, the Rank MSE models correspond to a $2.5\times$ improvement in the normed rank and a $1.8\times$ improvement with respect to the selection error.

The Rank MSE models, on average, predict layouts that achieve about 89\% of the Hellinger fidelity of the optimal layout (i.e., the selection error is about 11\%).  
The median normed rank predicted by these models is about 0.13, and the top layout is selected 22\% of the time.  
In comparison, Mapomatic achieves a selection error of about 19\%, an average normed rank of 0.34, and predicts the top layout 15\% of the time. 

We can also compare layout selection methods using the win rate
\begin{equation}
    \label{eq:winrate}
    \text{win rate} = \frac{\text{wins}+ \frac{1}{2} \times \text{ties}}{\text{total games}} \,,
\end{equation}
which treats each layout selection problem as a ``game'', and assigns a win to the method that predicted a better-performing layout.  Ties are reserved for the case when both methods predict equally performant layouts.
A 50\% win rate of a method compared to the baseline signifies an equivalence between the methods, while a win rate above (below) 50\% signifies superiority (inferiority) compared to the baseline.  
\footnote{Given that the empirical Hellinger fidelity will be subjected to shot noise, it would be sensible to define wins and ties within some tolerance. However, averaging over a dataset of many circuits will reduce sensitivity to these fluctuations.}
The baseline ranking method (Mapomatic) outperforms random selection, as indicated by the 35\% win rate listed in Table~\ref{table:results} (equivalently, Mapomatic ``wins'' against random selection 65\% of the time).
The learned circuit score models all consistently outperform Mapomatic, as indicated by win rates >50\%.
In particular, our best models exhibit a win rate of ~67\%, indicating a substantial improvement over the baseline method.

\section{Discussion\label{sec:discussion}}
In this work, we considered the general problem of identifying the best device or sub-system of a device and focused on a particular instantiation of this problem in the form of layout selection for a single device.
Layout selection is an oft-overlooked component of quantum compilation and transpilation.  
Omitting this step by simply selecting a layout randomly is strongly sub-optimal.

We developed a novel score-based method for ranking quantum circuit performance as a form of predictive QCVV. The score function is trained by minimizing a loss function that penalizes deviations from the ideal ranking over a large dataset of circuits containing a wide range of contextual errors that are likely to appear in circuits of interest, and which are missed through an exclusive reliance on simplified proxy measures.  
Once trained, the score function is efficient to compute and introduces no additional overhead.   
We experimentally verified on IBM hardware that our method provides a $1.8\times$ improvement in selection error when compared to a baseline approach, and $3.2\times$ when compared to random layout selection.

The results we present are generically applicable to ranking problems in the execution of hardware circuits and can be adapted beyond layout selection within a single device.  For instance, as an increasing number of hardware providers expose their devices to the public, identifying the best device for a given task is an important capability. Various algorithmic benchmarking routines have been explored to enable comparison between quantum processors, but linking benchmarking datasets to prediction for a user's specific algorithm has not been possible \cite{lubinski2023application, tomesh2022supermarq}.
The learning-to-rank protocol developed here can be directly applied to this problem, potentially even leveraging training data derived from the various benchmarking packages in use.

\subsection*{Data availability}
The data supporting the findings of this study are publicly available in Zenodo at \url{https://doi.org/10.5281/zenodo.14037434} \cite{hartnett2024learningdata}. \\

\subsection*{Acknowledgments}
The authors are grateful to all other colleagues at Q-CTRL whose technical, product engineering, and design work has supported the results presented in this paper.

\appendix
\onecolumn

\section{Phenomenological circuit score \label{app:score}}
The overall circuit score is expressed as a product of circuit scores for four mechanisms, $S = \prod_a (S_a)^{p_a}$, where $a$ runs over gate, measurement, $T_1$, and $ZZ$ errors. 
The power parameters $p_a$ are introduced to account for possible systematic under/over-weighting of these different errors.
Explicitly,
\begin{equation}
     S = (S_{\text{gate}}{}^{p_{\text{gate}}}) \, (S_{\text{msmt}}{}^{p_{\text{msmt}}}) \, (S_{T_1}{}^{p_{T_1}}) \, (S_{ZZ}{}^{p_{ZZ}}) \,.
\end{equation}

Gate errors are modeled as independent binary events, with each gate $g$ inducing an error with probability ${1-f(g)}$, where $f(g)$ is the fidelity of that particular gate. 
Note that here the gates are taken to be specified to the qubits upon which they act - for example, an $X$ gate acting on qubit $1$ has, in general, a different fidelity than an $X$ gate acting on qubit $2$. 
Thus, the gate score for circuit $C$, as an estimate of the total gate fidelity, is modeled as:
\begin{equation}
    S_{\text{gate}}(C) = \prod_g f(g)^{n(C; g)} \,,
\end{equation}
where the product is over all native gates for the device in question, and $f(g)$, $n(C; g)$ are the fidelity and count of gate $g$, respectively. 
The gate fidelities are device constants that can be obtained directly from the backend, typically through Randomized or Cross-Entropy Benchmarking \cite{emerson2005scalable, knill2008randomized, boixo2018characterizing, neill2018blueprint}, which provides a single fidelity measure that is insensitive to the circuit context.
However, in this work, we treat $f(g)$ and $f(\text{msmt}_i)$ as learnable parameters initialized to the estimated values obtained from the backend.

The measurement error is modeled similarly,
\begin{equation}
    S_{\text{msmt}}(C) = \prod_{i=0}^{n-1} f(\text{msmt}_i)^{n(C; \text{msmt}_i)} \,,
\end{equation}
where msmt$_i$ represents the measurement operation on qubit $i$, and $f(\text{msmt}_i)$ corresponds to one minus the mean miss-classification rate.
Note that measurement error mitigation has been applied to the executed circuits, effectively reducing this error rate.

Next, we model the effect of $T_1$ decoherence on idle qubits. 
According to the Bloch-Redfield decoherence model \cite{krantz2019quantum}, the time-dependent density matrix of an idle qubit in a general pure state $|\psi\rangle = \alpha |0\rangle + \beta |1\rangle$ evolves as:
\begin{equation}
    \rho(t) = 
    \begin{pmatrix}
    1 + (|\alpha|^2 - 1)e^{-\Gamma_1 t} & \alpha \beta^* e^{-\Gamma_2 t} \\ 
    \alpha^* \beta  e^{-\Gamma_2 t} & |\beta|^2 e^{-\Gamma_1}
    \end{pmatrix} \,,
\end{equation}
where $\Gamma_1 = 1/T_1$ is the longitudinal relaxation rate, $\Gamma_2 = \Gamma_1/2 + \Gamma_{\phi}$ is the transverse relaxation rate, which includes a contribution from the longitudinal rate as well as a pure dephasing term $\Gamma_{\phi} = 1/T_2$. 
The fidelity between the time-dependent and initial states is:
\begin{align}
    F( \rho(0), \rho(t) ) = |\alpha|^2 &+ \left(|\alpha|^4 + |\beta|^4 - |\alpha|^2\right) e^{-\Gamma_1 t} + 2 |\alpha|^2 |\beta|^2 e^{-\Gamma_2 t} \,, 
\end{align}
This fidelity is clearly state-dependent. 
It decays exponentially to the minimal value of $|\alpha|^2$, with $|\alpha| = 1$ corresponding to the special case where the initial state is at the North Pole of the Bloch sphere and is therefore unaffected by these decoherence effects. 

For the purpose of developing a simple phenomenological error metric applicable to typical quantum circuits, we average over all possible input states using the Haar measure.
As our circuit implementation includes dynamical decoupling, the effect of $T_2$ is highly suppressed. To isolate the effect of $T_1$ we set $\Gamma_{\phi} = 0$. %
The final result is that the fidelity of an average single-qubit state subject to $T_1$ decoherence for an idle time $t$ is
\begin{align}
    f_{T_1}(t; \Gamma_1, a, b) :&= \bar{F}( \rho(0), \rho(t) ) = (1 - a - b) + a \, e^{-\Gamma_1 t/2} + b \, e^{-\Gamma_1 t} \,, 
\end{align}
where $a = 1/3$, $b=1/6$ are constants obtained from the Haar average. 
In the machine learning model, $a$ and $b$ are promoted to model parameters which are initialized to the Haar-averaged values.
Note that the rate parameter $\Gamma_1$ will be both device- and qubit-dependent. We use $\Gamma_1$ the values provided by the backend during the circuit execution.

To obtain a circuit-wide fidelity metric, we assume that the $T_1$ errors act independently for each qubit and for each idle period: 
\begin{equation}
    S_{T_1} = \prod_{i = 0}^{n-1} \prod_{t \, \in \, \text{idle intervals(i)}} f_{T_1}(t; \Gamma_1^{(i)}, a, b) \,.
\end{equation}
Here $i$ runs over each qubit in the device, and the notation idle intervals$(i)$ is used to indicate the set of all idle periods for the $i$-th qubit. 
Notice that $S_{T_1}$ accounts for decoherent errors during idle periods only. Decoherent errors during gate operations are captured by $S_{gate}$.

The next error we account for is coherent error due to $ZZ$ crosstalk for connected and mutually idle qubits, and is modeled via $S_{ZZ}$. In cases where qubits are active, the $ZZ$ interaction with their neighbors is suppressed due to the inherent echo structure of the driving pulses.

The $ZZ$ interaction term for a pair of qubits $i$, $j$ is $H_{ZZ} = (\omega_{ZZ})_{ij} Z_i Z_j / 2$. 
In terms of the computational basis states for the $i$, $j$ qubits, the effect of the interaction is to induce an overall phase shift, with the sign of the phase dependent on the parity of the state: 
\begin{equation}
    |\psi_e \rangle \rightarrow e^{-i (\omega_{ZZ})_{ij} t/2} |\psi_e \rangle \,, \qquad |\psi_o \rangle \rightarrow e^{i (\omega_{ZZ})_{ij} t/2} |\psi_o \rangle \,,
\end{equation}
where $|\psi_e\rangle$ represents an arbitrary linear combination of the even-parity states $|00\rangle, |11\rangle$, and $|\psi_o\rangle$ represents an arbitrary linear combination of odd-parity states $|01\rangle$, $|10\rangle$. 
$ZZ$ coherent errors can be suppressed by applying context-aware dynamical decoupling (DD) \cite{mundada2023experimental}. While a perfect timing of DD sequence can eliminate the error completely, a sub-optimal timing leads to a mismatch duration for which the effect of $ZZ$ crosstalk is not canceled. 

As with the $T_1$ error, the $ZZ$ error is state-dependent, and therefore we perform a Haar average to obtain a general expression that can be used to estimate the $ZZ$ fidelity for arbitrary idle blocks within a given circuit: 
\begin{equation}
    f_{ZZ}(\Delta t; (\omega_{ZZ})_{ij}, c) = 1 - c \, \sin^2\left( \frac{(\omega_{ZZ})_{ij} \Delta t}{2} \right) \,,
\end{equation}
In the above expression, $\Delta t$ represent the $ZZ$ mismatch duration in a given mutual idling period. In the absence of DD, the mismatch duration is identical to the full idling duration while in the presence of an optimal DD the mismatch duration would go to zero.

Haar-averaging corresponds to setting $c=2/3$; however, as with the $a, b$ constants in the $T_1$ score, $c$ will be promoted to a learnable parameter in the machine learning model.
As before, to obtain a circuit-wide fidelity metric, we assume that the errors act independently for each pair of connected qubits and for each separate mutual idle period: 
\begin{equation}
    S_{ZZ} = \prod_{(i,j) \in G} \prod_{\Delta t \, \in \, \text{idle intervals$(i,j)$}} f_{ZZ}(\Delta t; (\omega_{ZZ})_{ij}, c) \,.    
\end{equation}
Here, $G$ is the device coupling graph (undirected), and the outer product is over pairs of connected qubits. The inner product is over periods where both qubits in question are idle. 

The final subset of parameters to consider are the powers $p_a$, which are introduced to give the model additional flexibility.
There is a redundancy associated with these, as the ranking induced by $S$ is invariant under rescaling by a positive constant, i.e. 
\begin{equation}
    (p_{\text{gate}}, p_{\text{msmt}}, p_{T_1}, p_{ZZ}) \rightarrow \lambda \, (p_{\text{gate}}, p_{\text{msmt}}, p_{T_1}, p_{ZZ}) \,, 
\end{equation}
with $\lambda > 0$. 
This motivates the identification of two sets of powers if they differ by an overall positive scale.
This may be accomplished by parameterizing the powers in a manner that allows the overall scale to be fixed; thereby removing the redundancy. 
A natural choice is to introduce spherical coordinates over the $\mathbb{R}^4$ spanned by the power parameters $p_a$ and to fix the radius to be 1, which has the effect of restricting the parameters to the unit $S^3$.
For example, using Hopf coordinates yields
\begin{subequations}
\begin{equation}
    p_{\text{gate}} = \cos \xi_1 \sin \eta \,,
\end{equation}
\begin{equation}
    p_{\text{msmt}} = \sin \xi_1 \sin \eta \,,
\end{equation}
\begin{equation}
    p_{T_1} = \cos \xi_2 \cos \eta \,,
\end{equation}
\begin{equation}
    p_{ZZ} = \sin \xi_2 \cos \eta \,.
\end{equation}
\end{subequations}
Technically, we are only interested in the region of $S^3$ where all powers are positive, which restricts the angular ranges to $\xi_{1,2} \in [0,\pi/2]$, $\eta \in [0, \pi/2]$ (the full sphere may be parameterized using $\xi_i \in [0, 2\pi]$). 

With the exception of the power $p_a$ and the Haar-averaged constants $a, b, c$, the parameters appearing in the score function are related to device physics and can be estimated via calibration routines. 
In our machine learning approach, these are promoted to be learnable parameters that are fit by minimizing a loss function over a dataset, Eq.~\ref{eq:loss}.
However, an important point is that the device-specific parameters naturally vary with time (hence the need to routinely re-estimate them).
To account for the fact that the dataset consists of circuit data executed during a period lasting several months, we implemented a parameterization of the circuit score that differs slightly from the above presentation.
Parameters that we expect to vary over time, such $f(g)$ (the gate fidelity for gate $g$), were re-parameterized as $f(g) = f_t(g)^{\lambda(g)}$, where $f_t(g)$ is the time-dependent value as reported by the backend, and where $\lambda$ is the parameter fit in the training step.

Lastly, for reference, the Mapomatic score function is:
\begin{equation}
    \label{eq:mapomatic}
    S_{\text{Mapomatic}}(C) = \prod_g f(g)^{n(C; g)} \, \prod_{i=0}^{n-1} f(\text{msmt}_i)^{n(C; \text{msmt}_i)} \,,
\end{equation}
where the first product runs over all native gates $g$ and the second product runs over the qubits.
Here $f(o)$, $n(C; o)$ are the fidelity and count of the operation $o$, respectively, and $f(\text{msmt}_i)$ corresponds to one minus the mean miss-classification rate for qubit $i$.
In this expression, both the gate and measurement fidelities are automatically estimated by IBM several times a day using common QCVV methods and obtained directly from the backend.

\section{Loss functions \label{app:lossfunctions}}
Given a parametric circuit score $S$, the learning task is to perform an optimization of the model parameters $\theta$ so that the ranking induced by the score closely approximates the empirical ranking provided by the Hellinger fidelity. This is an example of a Learning to Rank problem, which has been well-studied in the machine learning literature \cite{liu2009learning}. The basic strategy is to introduce a loss function $\mathcal{L}(\theta)$ that quantifies the extent to which the ranking induced by the circuit score $S$ differs from the ``true'' ranking, which here we take to be given by the Hellinger fidelity $H_F$. Then, this loss function is averaged over a training dataset $\mathcal{D}_{\text{train}}$, and then minimized with respect to the model parameters:
\begin{equation}
    \label{eq:loss}
    \theta_* = \text{argmin}_{\theta} \, \mathcal{L}(\theta) \,.
\end{equation}
Here, $\theta$ schematically refers to the full vector of model parameters. We considered multiple loss functions for training the circuit score (summarized in Table~\ref{table:lossfunctions}). These can be categorized according to the nature of the comparisons being made. Pointwise losses do not involve any comparisons among different layouts (i.e., the loss is computed on a per-layout basis), pairwise losses involve comparisons between two layouts, and listwise losses involve comparisons across the set of all layouts for a given initial circuit.

The sole pointwise loss function considered here is the Mean Squared Error (MSE) between the empirical Hellinger fidelity and the circuit score:
\begin{equation}
    \mathcal{L}_{\text{Score MSE}}(\theta) = \frac{1}{\left| \mathcal{D}_{\text{train}} \right|} \sum_{C \in \mathcal{D}_{\text{train}}} \left( S(C) - H_F(C) \right)^2 \,. 
\end{equation}
This loss encourages the circuit score to match the fidelity. While straightforward, this score MSE loss might be too strong of an objective, given that the goal is for the score to merely reproduce the ordinal ranking induced by the Hellinger fidelity, as opposed to the precise numerical value for each circuit. In particular, any monotonic transformation of the circuit score will leave the ranking unchanged. This motivates loss functions that compare two or more layouts. 

A common pairwise approach is to use the binary cross entropy (BCE) loss:
\begin{equation}
    \mathcal{L}_{\text{Pairwise BCE}}(\theta) = - \frac{1}{2} \sum_{\substack{C_1, C_2 \in \mathcal{D}_{\text{train}} \\ : C_1 \neq C_2}} 
    [[H_F(C_i) < H_F(C_j)]] \ln P_{ij} \,. 
\end{equation}
Here $[[x]]$ is the Iverson bracket (evaluating to 1 if $x$ is true and 0 otherwise), and $P_{ij}$ is the probability under the model that the circuit $j$ has a higher Hellinger fidelity than circuit $i$. This is commonly modeled as the logistic sigmoid of the score difference \cite{burges2005learning},
\begin{equation}
    P_{ij} = \frac{1}{1 + \exp\left( -(S(C_i) - S(C_j)) \right) } \,.
\end{equation}
As discussed in the main text, the pairwise loss necessitates a quadratic expansion of the dataset: an initial dataset of $N$ distinct circuits yields $N(N-1)/2$ distinct pairs of circuits. This expansion can be useful because it allows for many more training examples to be drawn from a given dataset, but it also has drawbacks because the dataset size can become unwieldy as well as depart from being independent and identically distributed (iid) \cite{cao2007learning}. For these reasons, we did not implement a pairwise loss in our experiments.

The listwise approach to Learning to Rank assigns a scalar loss to an entire set of circuits. This is naturally suited to layout selection given that the problem is to correctly predict the best performing circuit layout out of the set of all possible layouts. In these approaches, the dataset is divided into $N_B$ batches. Each batch consists of the set of layouts derived from a single initial circuit. A listwise approach will then assign a loss value to each batch characterizing the extent to which the rankings induced by the Hellinger fidelity and circuit score differ. 

We considered a number of listwise loss functions. The Pearson correlation loss is motivated by the observation that an ideal circuit score will be perfectly correlated with the Hellinger fidelity, and therefore the negative Pearson correlation may be used as a loss function:
\begin{equation}
    \mathcal{L}_{\text{Pearson}}(\theta) = -\frac{1}{N_B} \sum_{i=1}^{N_B} r_{H_F, S} \,.
\end{equation}
Here $r_{x,y} \in [-1,1]$ is the Pearson correlation coefficient for two sequences $x, y$, with $r_{x,y} = 1$ corresponding to perfect correlation (sequence indices have been suppressed). Note that the Pearson correlation coefficient is a biased estimator of the true correlation $\rho(X,Y)$.

The closely related Spearman correlation loss is motivated by the observation that, although two monotonically related circuit scores produce the same ranking, the Pearson correlation loss is not invariant under monotonic transformations of the circuit score. The Spearman correlation loss addresses this by replacing the Pearson correlation coefficient with the Spearman rank correlation coefficient $r^{(s)}_{x,y} = r_{R(x), R(y)}$, which is just the Pearson correlation of the ranks of each variable $R(x)$, $R(y)$. This loss is manifestly invariant under monotonic transformations of the circuit score due to the invariance of the rank function. An important technical obstacle to using the Spearman loss is the fact that the rank function is not differentiable, which is a necessary condition for gradient-based optimization algorithms. We surmount this obstacle by computing a ``soft'' Spearman rank correlation coefficient using the method of \cite{blondel2020fast}, which introduced a relaxed definition of the rank function based on projections onto the permutahedron (the convex hull of permutations). This prescription allows the Spearman correlation loss to be optimized using first-order approaches such as gradient descent.

\begin{figure*}[htpb!]
    \centering
\includegraphics[width=1.0\textwidth]{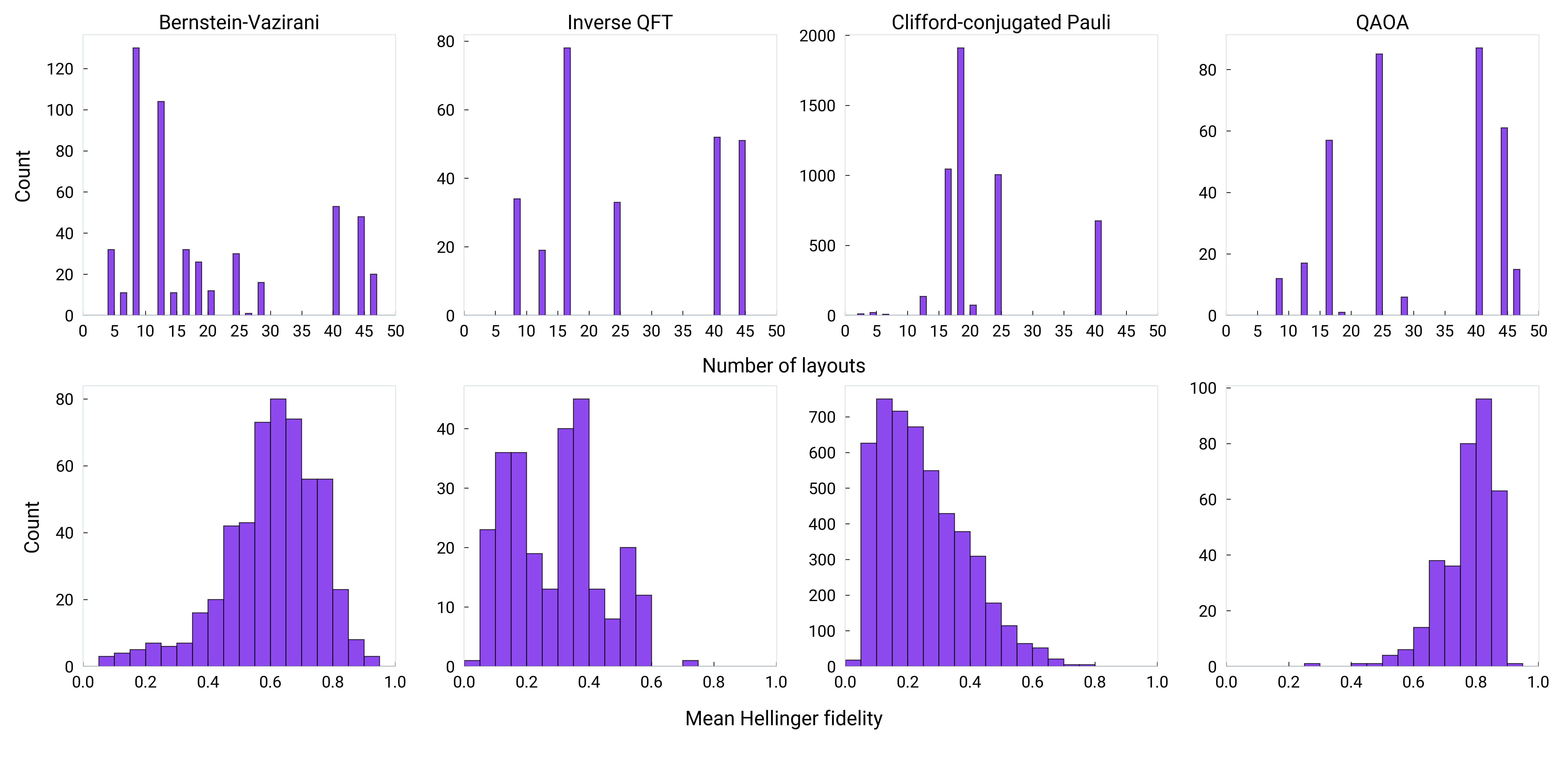}
    \caption{
    Dataset statistics. (\textit{Top}): The distribution of layout sizes (the number of ways that a given initial circuit can ``fit'' onto the device after hardware transpilation), according to algorithm. The number of layouts per circuit ranges from 2 to 46. (\textit{Bottom}): The distribution of empirical Hellinger fidelities, also according to algorithm. Note that the details of each algorithm, such as the circuit width $n$ or the random graph used in the QAOA circuit were chosen to produce circuits whose Hellinger fidelities were not extreme (i.e., near 0 or near 1).
    }
    \label{fig:dataset_statistics}
\end{figure*}

Next, the rank MSE loss function is motivated by the fact that the Spearman rank correlation coefficient between two length-$L$ sequences $x$ and $y$ can be written in terms of the squared difference of ranks:
\begin{equation}
    r^{(s)}_{x,y} = 1 - \frac{6 \sum_{i=1}^L \left( R(x_i) - R(y_i) \right)^2}{L(L^2 -1)} \,.
\end{equation}
Consequently, minimizing the Spearman loss is equivalent to minimizing the MSE of the rank. In many ranking problems, including layout selection, errors among the top-ranked items should be more costly than errors among low-ranked items, which suggests a simple modification of the above:
\begin{equation}
    \label{eq:rankMSE}
    \mathcal{L}_{\text{Rank MSE}}(\theta; d) = \frac{1}{N_B} \sum_{i=1}^{N_B} \sum_{\ell=1}^{L_i} \frac{\left( R(H_F(C_{\ell})) - R(S(C_{\ell}) \right)^2}{ R(H_F(C_{\ell})) ^d} \,,
\end{equation}
where $d$ is a hyper-parameter used to control the extent to which top-rank errors should be over-penalized relative to low-rank errors. The rank MSE loss with $d=0$ is equivalent to the Spearman loss in that the two loss functions are the same up to linear transformation.

Lastly, another way to bias the loss function towards the top-ranked layouts is to maximize the probability of observing the empirically-observed top-$K$ layouts across all batches, ${\prod_{i=1}^{N_B} \text{Pr}( \text{top } K)_i}$. This is equivalent to minimizing the negative log-likelihood:
\begin{equation}
    \mathcal{L}_{\text{NLL}}(\theta; K) = - \frac{1}{N_B} \sum_{i=1}^{N_B} \ln \text{Pr}( \text{top } K)_i \,. 
\end{equation}
An expression for $\text{Pr}( \text{top } K)_i$ is required to implement this loss. We adopt the Plackett-Luce model, which models this probability as a product of normalized scores \cite{plackett1975analysis}:
\begin{equation}
    \text{Pr}( \text{top } K) = \prod_{k = 1}^K \frac{ S_{j_k} }{ \sum_{\ell \in A_k} S_{\ell}} \,.
\end{equation}
Here $j_k$ is the index of the circuit with the $k$-th empirical Hellinger fidelity and ${A_k = A_{k-1} \setminus \{j_k\}}$ with ${A_1 = \{1, 2, ..., L\}}$ is the set of alternatives. To explain the notation with an example, suppose that for a circuit with $L=4$ layouts $\{C_1, C_2, C_3, C_4\}$ the following ranking of empirical Hellinger fidelities is observed: $H_F(C_2) > H_F(C_4) > H_F(C_1) > H_F(C_3)$. Then
\begin{equation}
    \text{Pr}( \text{top } 3) = 
    \left( \frac{S(C_2)}{\sum_{\ell \in \{1,2,3,4\}} S(C_{\ell})} \right) \left( \frac{S(C_4)}{\sum_{\ell \in \{1,3,4\}} S(C_{\ell})} \right) \left( \frac{S(C_1)}{\sum_{\ell \in \{1,3\}} S(C_{\ell})} \right) \,.
\end{equation}

\section{Circuit dataset \label{app:additionalresults}}
The circuit ensemble was designed with the intention of being representative of the types of circuits that users of near-term quantum devices are interested in running and encompassing a diverse set of circuit contexts.
For this approach to scale to larger devices, an additional criterion is that the circuits should have a known output or should be efficiently simulable, so that the ideal bitstring distribution, and hence the Hellinger fidelity, may be classically computed.
This was accomplished by restricting attention (with one exception) to circuits whose ideal, non-noisy output state is guaranteed to be a single computational basis state, i.e. $U(C)|0\rangle = |i\rangle$, $i \in \{0,1,...,2^n-1\}$, with $U(C)$ the unitary implemented by the circuit $C$. 
Such ``one-hot'' or deterministic circuits are appealing because the Hellinger fidelity corresponds to the circuit success probability. 

The Bernstein-Vazirani, Inverse QFT, and Clifford-conjugated Pauli circuits all satisfy the above criteria.
The dataset was augmented with a fourth algorithm, QAOA \cite{farhi2014quantum}, that does not satisfy these criteria.
This choice was made to improve the relevance of the dataset to circuits of interest to users of near-term noisy devices today.
However, it is important to note that the QAOA circuits are not efficiently simulable, and therefore this augmentation approach will not scale to larger devices.

As described in the main text, we generated such a dataset for the \textit{ibmq\_guadalupe} device, consisting of 6020 initial circuits, 81\% of which are the Clifford-conjugated Pauli circuits, 9\% Bernstein-Vazirani, 6\% QAOA, and 4\% inverse QFT. 
After accounting for all allowed layouts, these 6020 initial circuits became 132,866 circuits after hardware transpilation. 
These specific proportions are chosen to ensure that a wide diversity of circuit contexts is present in the dataset and to avoid identical or close-to-identical circuits in the data set. 
Highly structured algorithms such as BV or QFT transpile to different circuits as the qubit count $n$ is varied or due to the variations in the transpilation procedure. 
However, given that the dataset corresponds to circuits executed on a 16-qubit device, the number of distinct transpiled circuits for these structured algorithms is highly limited. 
Simply put, it is not possible to obtain a large set of unique transpiled BV circuits for $n\le 16$.
Consequently, the dataset is dominated by random circuits which can naturally generate a greater diversity of circuit contexts.
Moving forward, larger and better-quality devices should allow for greater flexibility in designing the circuit ensemble.

Lastly, in Fig.~\ref{fig:dataset_statistics} we show the distribution of layouts (per initial circuit) according to algorithm, as well as the distribution of mean Hellinger fidelities for each initial circuit.

\bibliography{refs}
\bibliographystyle{quantum}

\end{document}